# Phenalenyls as tunable excellent molecular conductors and switchable spin filters

Manuel Smeu,[a]* Oliver L.A. Monti,[b,c] and Dominic McGrath[b]

[a] Department of Physics, Binghamton University – SUNY
[b] Department of Chemistry and Biochemistry, University of Arizona
[c] Department of Physics, University of Arizona

**ABSTRACT:** We demonstrate a new class of molecules for exceptional performance in molecular electronics and spintronics. Phenalenyl-based radicals are stable radicals whose electronic properties can be tuned readily by heteroatom substitution. We employ density functional theory-based non-equilibrium Green's function calculations to show that this class of molecules exhibits tunable spin- and charge-transport properties in single molecule junctions. Our simulations identify the design principles and interplay between unusually high conductivity and strong spin-filtering: Paired with moderate conductance ($10^{-3}G_0$), two of the four radicals investigated exhibit above 80% spin filter efficiency that is moreover tunable via bias control. Conversely, two radicals that make modest spin filters are excellent conductors with a low bias conductance reaching $0.48G_0$. This is made possible by the unusually good alignment of the singly occupied or unoccupied molecular orbital with the Fermi level of the electrodes, overcoming the limitations of Fermi level pinning that typically plague molecular electronics. We show that this interplay between excellent conductance and high spin-filter efficiency is determined by the energy alignment between the singly (un)occupied molecular orbital and the Fermi level of the electrodes, and that for phenalenyls this can be controlled with judicious heteroatom substitution.

Creating pure spin current is a fundamental challenge in spintronics, with enormous potential for magnetic storage, low-power electronics, quantum information science, and a fundamental understanding of the interplay between spin and current.[1] Most commonly, spintronic devices such as magnetic tunnel junctions and spin valves make use of carefully engineered stacks of thin magnetic and non-magnetic materials to create *e.g.* spin-transfer torque structures.[1] Recently, the use of molecules for creating spin current has received considerable attention: *E.g.*, spin-polarized current may be achieved from transport through chiral molecules.[2,3] Alternatively, all-organic radicals may provide a means for achieving strong spin-filtering and long spin-coherence times at the nanoscale, without the need for external magnetic fields, ferromagnetic electrodes, or transition metal atoms. However, detailed investigation has been hampered by the limited number of stable all-organic radicals.[4–8]

Standard organic spin valve designs require ferromagnetic electrodes and low-temperature operation to achieve significant spin polarization.[9,10] In such devices, the degree of spin polarization is typically determined by the combination of molecule and electrodes, the "spinterface,"[11] and the spin current originates from the spin-polarized density of states in the electrodes. Alternatively, and without the need for ferromagnetic electrodes, an $S = \frac{1}{2}$ organic radical may support a spin-split density of states in the junction, leading to tunable differential transmission in the two spin channels.[12]

Here we propose that the single most important aspect of a single-molecule spin filter is the energy level alignment at the molecule-electrode interface, *i.e.*, the extent to which one of the molecular radical orbitals remains singly occupied. We showed previously that charge transfer from the electrodes to the molecule can result in the loss of electron spin polarization in transport, eliminating the spin-filter effect.[13] Here we show how the interplay of charge-transfer, energy level alignment and spin-polarized transport in a new class of stable all-organic radicals enables the design of single molecule junctions with exceptionally high conductance, or high-efficiency and tunable spin filtering.

Efforts to increase electron transmission (conductance) by molecular design have had limited success, attributed in part to Fermi level pinning,[14] an effect that creates a significant injection barrier that appears to be difficult to eliminate. This limitation hinders the realization of the true potential of single molecule electronics. Radials may offer a solution to this problem. In a spin-restricted picture, the relevant molecular orbital (MO) is half-filled only, and may be aligned with the Fermi level ($E_F$), conferring metallic character to the molecule,[15] which may offer a route to bypassing the Fermi level pinning problem. Allowing for the two spins to be different (unrestricted) does not modify this picture substantially, and rather small injection barriers may be realized,[16] suggesting that conductance may be increased by orders of magnitude compared to closed-shell molecules.[17–19]

In this communication, we investigate four structurally similar radicals belonging to the class of phenalenyl (PLY) molecules with tailored electronic and spin properties by heteroatom substitution in the ring system, illustrated in Chart 1. For this investigation we restrict ourselves to -SH anchor groups. PLY radicals are known to be stable towards dimerization in solution due to the delocalized spin density;[20,21] synthetic routes are extant for the PLY, 1,3-DAPLY, 4,9-DOPLY and 1,3-DA-4,9-DOPLY cores,[22–25] which proceed from readily available naphthalene derivatives. We employed density functional theory (DFT) to calculate the gas phase MO energies. The molecules were then relaxed between Au electrodes to form two-probe junctions, and we employed the non-equilibrium Green's function technique in conjunction with DFT (NEGF-DFT)[26] to investigate their spin-resolved electron transport properties, including electron transmission spectra, current-voltage characteristics, and spin filter efficiency.

**Chart 1.** The four phenalenyl class radicals and the closed-shell naphthalene analog.

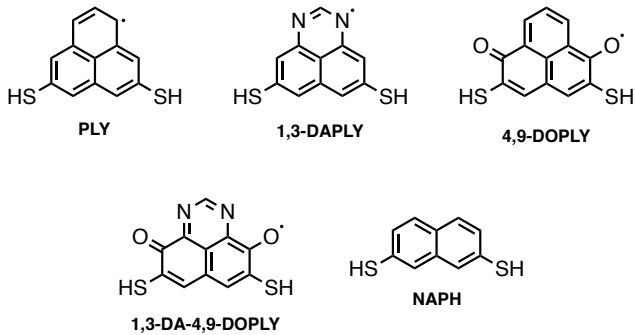

To begin, we calculated the frontier MOs of the four PLY-based radicals in the gas phase. These consist of the singly occupied molecular orbital (SOMO; $\alpha$ spin) and the singly unoccupied molecular orbital (SUMO; $\beta$ spin), with their corresponding energies in eV, as calculated with the Orca code[27] at the B3LYP/6-311++G(d,p) level of theory[28,29] (Figure 1). We note that there is a substantial SOMO-SUMO gap in the radicals that ranges from 1.5 to nearly 2 eV. While the SOMO and SUMO have the same orbital structure for each radical, the MOs differ greatly across the different radicals, with 4,9-DOPLY and 1,3-DA-4,9-DOPLY supporting substantial amplitude at the anchoring group, while PLY, and 1,3-DAPLY have nodes at the -SH linker group.

The spin-resolved transmission functions [$T(E)$] were calculated with NEGF-DFT (Figure 1) as described in the SI. As expected, the $\alpha$ and $\beta$ transmission spectra are identical for the closed-shell NAPH system. Conversely, all four radicals exhibit spin-polarized transport as can be seen from the spin-split $\alpha$ and $\beta$ transmission spectra (blue and red plots), particularly near $E_F$. Among the four radicals, all have the SOMO transmission peak just below $E_F$, and the SUMO peak above $E_F$, as shown by the blue and red arrows in Figure 1 (the transmission peaks are assigned in Section S2 of the SI). While the heteroatom substitution achieves a large difference in SOMO energy of the isolated radicals (left of Figure 1), this only manifests as a small difference in the SOMO transmission peak positions in the junction for the four radicals, aligned just below $E_F$. In contrast, the SUMO peak positions differ, resulting in a varying SOMO-SUMO transmission peak separation across the four radicals. This seemingly subtle point—the position of the SUMO transmission peak relative to the SOMO peak and $E_F$—is the determining factor that governs the ultimate charge/spin transport characteristics of these radicals, as discussed below.

Also of interest, 4,9-DOPLY and 1,3-DA-4,9-DOPLY have relatively broad SOMO/SUMO peaks, likely resulting from the MO amplitude on the -SH anchoring groups providing strong coupling to the electrodes, unlike PLY and 1,3-DAPLY which have MO nodes at the anchoring group, resulting in weaker coupling and much narrower transmission peaks (see Figure 1). Therefore, despite some systems (e.g., 1,3-DAPLY and 4,9-DOPLY) having very similar SOMO energies, they have completely different wavefunctions by simple heteroatom substitution/addition. This manifests both in different electronic coupling to the electrodes (transmission peak width) and ultimately different spintronic behavior. The zero-bias results presented in Figure 1 suggest that all radicals would result in spin-polarized electron transport and high conductance, based on the proximity of the SOMO peak to $E_F$. To further investigate these systems, we carried out bias-dependent NEGF-DFT calculations (see SI for details).

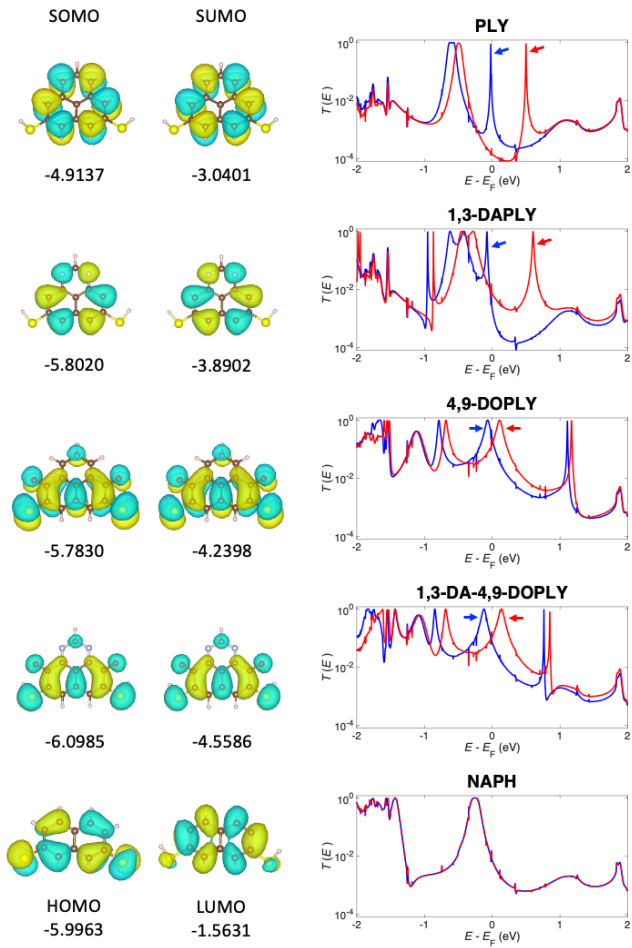

**Figure 1.** Left: Frontier molecular orbitals for the four radicals (SOMO/SUMO) and the naphthalene molecule (HOMO/LUMO); additional MOs are provided in Figures S2-S6 of the SI. The energies (in eV) are listed below each orbital. Right: Spin-resolved transmission coefficient [$T(E)$] for each radical/molecule bridging Au electrodes. Blue/red is for $\alpha/\beta$ electron transport; the arrows label the SOMO (blue) and SUMO (red) transmission peaks.

We consider the bias-dependent current ($I$-$V$) for the four radicals and naphthalene (Figure 2a). We note that, while all radicals have transmission peaks near $E_F$ (Figure 1), considerably more current flows through 4,9-DOPLY and 1,3-DA-4,9-DOPLY. To understand this result, we turn to the bias-dependent transmission spectra for 4,9-DOPLY (Figure 3a); both the SOMO and SUMO transmission peaks are included in the bias window (grey shaded region) at 0.2 V, which is possible due to the relatively low energy of the SUMO level for this radical. Since the SOMO and SUMO result in broad transmission peaks (plotted on a linear scale) to begin with, they yield high transmission (current) through this system. In contrast, the bias-dependent transmission spectra for PLY (Figure 3b) show a larger separation between the SOMO and SUMO peaks, which are also narrower (plotted on a logarithmic scale), resulting in comparatively smaller conductance through this radical.

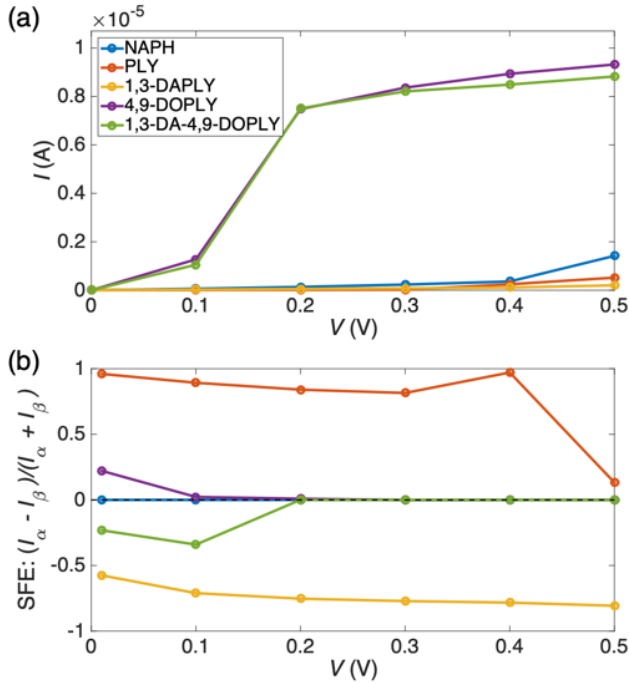

Figure 2. Bias-dependent current (a) and SFE (b) for naphthalene and the four PLY radicals.

It is apparent that both appropriate energy level alignment and coupling to the electrodes are needed to achieve the partial filling of the two spin-orbitals, resulting in the near-metallic conductance for (1,3-DA-)4,9-DOPLY. At a modest bias of 0.2 V, conductance reaches $0.48G_0$, an exceptionally high value for a single molecule coupled to Au electrodes with thiol anchoring groups; even the much smaller system of benzene dithiol exhibits a low-bias conductance of approximately $10^{-2}G_0$.[30]

We now turn our attention to spin-transport, and calculate the spin filter efficiency as

$$\text{SFE} = \frac{I_\alpha - I_\beta}{I_\alpha + I_\beta},$$

where $\alpha$ and $\beta$ represent spin-up and spin-down channels. SFE is plotted as a function of bias in Figure 2b. Remarkably, while all four radicals support some degree of spin-polarized current at low bias (0.01 V), only PLY and 1,3-DAPLY maintain a large magnitude of SFE up to 0.4 V; both 4,9-DOPLY and 1,3-DA-4,9-DOPLY entirely lose their SFE by the relatively modest bias of 0.2 V.

The SFE loss behavior of (1,3-DA-)4,9-DOPLY can be understood by inspecting the bias-dependent transmission spectrum. As an illustrative example, consider the bias-dependent transmission for 4,9-DOPLY (Figure 3a). The $\alpha$ and $\beta$ transmission spectra start out different at 0.0 V, but with increasing voltage and expanding bias window (gray shaded region in Figure 3), the SUMO peak (red) is pulled into the bias window, while the bias window is "catching up" to the SOMO peak. At 0.2 V, both the SOMO and SUMO peaks are contained within the bias window, setting the occupation of each level to a similar value, thereby eliminating the energy splitting and rendering the $\alpha$ and $\beta$ energy levels and their resulting transmission spectra spin-degenerate. This highlights that a radical exhibiting spin-polarized transmission at low bias may lose spin polarization even at modest bias. It is important to stress that this conclusion would be entirely missed without non-equilibrium (finite bias) electron transport calculations.

As reported previously,[13] a low-energy SUMO can lead to the radical accepting an electron from the electrodes, resulting in both SOMO and SUMO becoming occupied, and establishing degeneracy of energy levels and their transmission peaks. This is consistent with a charge analysis showing a strong correlation between the charge on the in-junction molecule and the SUMO energy of the isolated molecule (see Figure S7 in the SI).

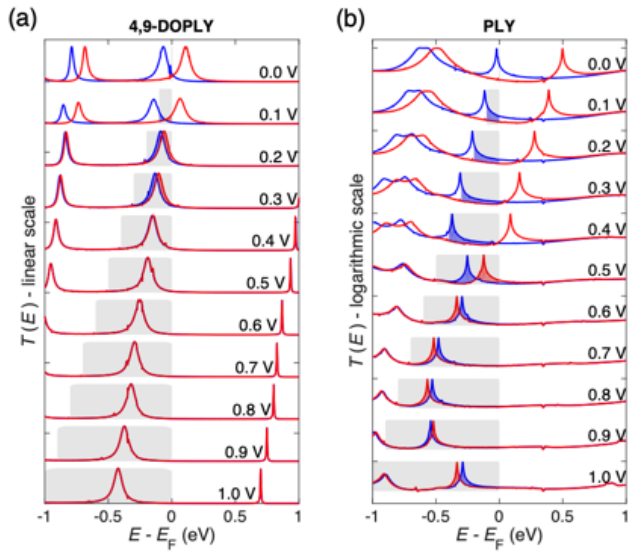

Figure 3. Bias-dependent and spin-resolved transmission spectra for (a) 4,9-DOPLY and (b) PLY radicals. The shaded gray regions enclose the bias window in which the transmission contributes to current. The shaded blue/red regions contain the excess $\alpha/\beta$ transmission at a given energy within the bias window. Note that (a) uses a linear scale while (b) uses a logarithmic scale.

In contrast, the two radicals with the larger SOMO-SUMO transmission peak separation maintain high SFE up to 0.4 V, with values near 0.9 for PLY and -0.8 for 1,3-DAPLY. The bias-dependent transmission spectra for PLY (Figure 3b) illustrate how the $\alpha$ and $\beta$ transmission spectra remain split, resulting in excess $\alpha$ transmission (shaded blue region) in a given bias window. However, at 0.5 V the SUMO peak enters the bias window, largely cancelling out the excess $\alpha$ spin transport, thereby reducing SFE. This is the first demonstration of bias-dependent tunable SFE in single molecule electronics. It can be turned on between 0 and 0.4 V, and turned off above 0.4 V, demonstrating tunable spin filtering over an experimentally accessible bias range of $\leq 0.5$ V.

The extent to which molecular spin filters are possible, and the factors that control the amount of spin polarization, have been rather unclear in the past; some computational investigations suggest that the spin density must be delocalized to achieve spin filtering and spin polarized current,[31] while others point towards the importance of the linker group.[32,33] In most of the PLY family radicals investigated, we find the spin density delocalized over most of the molecule (see Figures S2-S5 in the SI), yet $T(E)$ shows the tell-tale sign of strong spin polarization only in PLY and 1,3-DAPLY, suggesting that factors other than spin density are important. Our work suggests that the key factor that determines spin polarization is the energy level alignment: high SFE is achieved if only one of the spin-split radical levels is occupied. Alternatively, nearly complete occupation by an excess electron, *e.g.*, as with 4,9-DOPLY and 1,3-DA-4,9-DOPLY, collapses the spin-splitting in the transmission function, in exchange for highly-enhanced molecular conductance.

These classes of single molecule devices can be achieved with judicious heteroatom substitution, allowing for control over the frontier MOs (SOMO/SUMO), including their shape, affecting their coupling to the electrode states, and the ultimate charge and spin transport character of the molecule in-junction. Additionally, different anchoring groups and side groups can be envisaged to further tune electronic properties.

This work suggests that appropriate radicals such as the class of PLY can indeed move beyond Fermi level pinning and exhibit potentially record conductances that have so far only been achieved for special systems with covalent carbon-electrode bonds.[34,35] In the case of those radicals that maintain their radical character when bonded to Au electrodes, these results suggest the possibility for tunable spin filters that can be controlled by voltage. This effect may enable the creation of a switchable spin-diode in single molecules without the need of an external magnetic field or ferromagnetic electrodes.

## ASSOCIATED CONTENT

### Supporting Information

The Supporting Information is available free of charge at https://pubs.acs.org/XXX

Computational details of two-probe relaxations and NEGF-DFT calculations; larger set of MOs, spin density, $T(E)$ peak assignment; charge and spin polarization vs SOMO/SUMO energy; bias-dependent $I$-$V$ and SFE up to 1.0 V.

## AUTHOR INFORMATION

### Corresponding Author

* msmeu@binghamton.edu### Author Contributions

The manuscript was written through contributions of all authors. All authors have given approval to the final version of the manuscript.

### Funding Sources

OLAM and DVM gratefully acknowledge support by the National Science Foundation under award no. DMR-1708443.## ACKNOWLEDGMENT

This work was performed using the Spiedie High Performance Computing cluster at Binghamton University and the Expanse cluster, a part of the Extreme Science and Engineering Discovery Environment (XSEDE), which is supported by NSF Grant No. ACI-1548562 under allocation TG-DMR180009. We thank Dr. Gökhan Ersan for the TOC graphic. OLAM and DVM acknowledge support from the National Science Foundation (DMR-1708443).

Authors are required to submit a graphic entry for the Table of Contents (TOC) that, in conjunction with the manuscript title, should give the reader a representative idea of one of the following: A key structure, reaction, equation, concept, or theorem, etc., that is discussed in the manuscript. Consult the journal's Instructions for Authors for TOC graphic specifications.

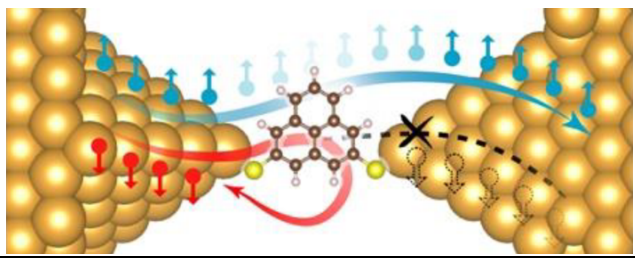

6